\documentclass[twocolumn,pre]{revtex4}
\usepackage{amsmath}
\usepackage{graphicx}
\usepackage{amssymb}
\usepackage{latexsym}
\usepackage{color}
\usepackage{multibox}

\makeatletter
\usepackage{epstopdf}
\def\P{\dot{\varepsilon}_P}

\def\scalardeformation{\varepsilon}

\makeatother
\begin{document}

    \title{An elastic, plastic, viscous  model for slow shear of a liquid foam}

\date{\today}

\author{Philippe  Marmottant }

\author{Fran\c cois Graner}

\affiliation{Laboratoire de Spectrom\'etrie
Physique, CNRS-Universit\'e Grenoble I,
B.P. 87, F-38402 St Martin d'H\`eres Cedex, France.}

\begin{abstract}
We suggest a scalar model for deformation and flow  of an
amorphous material such as a foam or an emulsion.
To describe elastic, plastic and viscous behaviours, we use
three scalar variables: elastic deformation, 
plastic deformation rate and
total deformation
rate; and three material specific parameters: shear modulus, yield
deformation and viscosity.
We obtain equations valid for different       types of  deformations and flows slower
than the relaxation rate towards mechanical equilibrium.
In particular, they are valid both in transient or steady flow regimes,
even at large elastic deformation. We
discuss why viscosity can be relevant even in this  slow shear
(often called ``quasi-static") limit.
Predictions of
the storage and loss moduli agree   with the
experimental literature, and explain with simple arguments the 
non-linear large amplitude trends. 

\end{abstract}

\maketitle

\section{Introduction}

Elastic
materials deform reversibly  \cite{lan86};
plastic  materials can be sculpted, that is, they
can be deformed into a new shape and keep it
\cite{cha87}; and  viscous  materials flow  \cite{bat00}.
A wide variety of materials display a combination of these properties, such as
elasto-plastic  metals and rocks, visco-elastic  polymer solutions
or visco-plastic  mineral suspensions \cite{mac94,fra93,fra95}.

Liquid foams, that is gas bubbles separated by liquid walls,  are
visco-elasto-plastic \cite{kra88,wea99,hoh05}: they are elastic
at low strain,
plastic at high strain and flow under high shear
rate.
This is also the case for other concentrated
suspensions of deformable objects in a liquid \cite{mac94,lar99,mas96},
such as  droplets
in emulsions, vesicles suspensions, or red blood cells
in blood.

Despite a large literature on experiments and simulations
(see \cite{hoh05} for a review),
we lack an unified theoretical description of foams.
There is no consensus yet on a central question:
what are the physically relevant variables?
A series of statistical models focus on fluctuations and their correlations
\cite{sol97,fal98,kab03,pic05,cou05}.
Conversely, recent contributions \cite{mar06,jan06,miy06,wys06,lab04,hoh06,sar06}
focus on average macroscopic quantities to obtain a more classical 
continuous description.

Here we choose to group three macroscopic quantities which are 
measurable as averages on
microscopical details \cite{mar06}:
(i) Elastic deformation  is a state variable \cite{por97} reversibly stored
by the foam's microstructure, that is,
the shape of bubbles \cite{aub03,asi03}; it determines the elastic 
contribution to the stress.
(ii) Plastic deformation
results in energy dissipation analogous
to solid friction.
  (iii) Large scale velocity gradients are associated with a viscous friction.
  Each of the three mechanical behaviors is associated with a material
specific parameter: elastic modulus, yield deformation and viscosity.

For simplicity, we assume here that these parameters are constant and
the equations are linear.
We consider here homogeneous deformation of a
material, not  depending on space coordinates.
We consider only
the magnitude of deformation, but not
spatial orientation: the material state variables are all
scalars.
This represents an incompressible liquid foam,
where the deformation is a pure shear.
We assume that this shear is slow enough so that the foam
is always close to mechanical equilibrium,
but quick enough to neglect coarsening such as due to gas diffusion
between bubbles, or bubble coalescence due to soap film breakage.
Although this model is minimal, it is written with enough generality
to enable for extensions to higher dimensions using tensors 
(the correspondance with tensors introduces a factor $1/2$, see section \ref{sec:Perspectives}),
 to higher shear rates,
and to other ingredients such as external forces (to be published).

This article is organised
as follows.
Section \ref{sec:qstat_model}   introduces a
visco-elasto-plastic  model
(eqs. \ref{eq-evol-u-scalar},\ref{eq:stresselasticviscous}) based on two scalar
variables: the elastic deformation and the (slow) shear rate (Fig. 
\ref{diagramme_vep}).
The rate of plastic deformation
is determined by  both the applied shear rate,
and the current state of the elastic  deformation (or equivalently
the  elastic part of the stress) rather than by the total stress
\cite{cri82,mil87}.
Section \ref{predict} presents scalar predictions of creep and
oscillatory responses. The storage and loss
moduli predicted   as a function of the strain amplitude agree with
experimental data without any adjustable parameters, using only the 
three model-independent parameters determined by experiments (yield 
point, shear modulus, viscosity).
The agreement becomes very good if we describe the plastic yielding
as a gradual transition spreading between an onset value of deformation and a 
saturation value   (eq.   \ref{eq-evol-u-smooth}).
Section \ref{disc} summarises and discusses our model, and opens some 
perspectives.

\section{Model}
\label{sec:qstat_model}

\subsection{Kinetics}

\subsubsection{Elastic and plastic strain}

The elastic deformation $U$ is a {\it state variable}, that is an 
intrinsic property of the foam's current deformation state. We note 
its time derivative $dU/dt$. Conversely, we use a dot for
 the total strain rate $\dot{\scalardeformation}$ and
the plastic strain rate $\P$, emphasising that they are not the time 
derivative of a state variable. For instance, the time integral
$\scalardeformation = \int  \dot{\scalardeformation}\; {\rm d}t$ of
the velocity gradient is the gradient of displacement (more 
generally, for large deformations, it is  a function of the 
displacement): it is extrinsic and {\it explicitly depends} on the 
sample's past history.

The total applied deformation rate is
shared between  elastic deformation $U$ and the
plastic deformation rate:
\begin{equation}
\dot{\scalardeformation} =\frac{{\rm d}U}{{\rm d}t} + \P.
\label{def_P}
\end{equation}

In the particular case of an elastic regime, $\P=0$,
    the elastic deformation $U$ is equal to the total applied deformation on
the material $\scalardeformation$.
Thus, in an elastic regime, no intrinsic definition of $U$ is necessary.

However, as soon as $\P \neq 0$, the situation changes.
  $U$ and $\dot{\scalardeformation}$ become independent variables,
and $\scalardeformation=\int \dot{\scalardeformation}dt$ does not 
define the elastic deformation.
In the extreme example of a steady flow, ${\rm d}U/{\rm d}t=0$,
then $\dot{\scalardeformation} =\P$: $U$ and $\dot{\scalardeformation}$ are no longer correlated.

These variables are macroscopic: $U$ is 
related to the elastic contribution to macroscopic stress and $\P$ to the  irreversibility 
of the stress {\it versus}  total strain curve.
In the specific case of foams, they can be traced back to detailed 
properties of the  bubbles pattern: independent, intrinsic 
definition  \cite{por97} based on geometry (shape of bubbles \cite{asi03}) for $U$;
and 
topological rearrangements
called ``T1 processes" \cite{lau02,gop03, kab03} 
(using their rate and orientation  \cite{mar06}) for $\P$.

\subsubsection{Sharing the total strain}

The problem now is to express how, in eq.  (\ref{def_P}),
$\dot{\scalardeformation}$ is shared between ${\rm d}U/{\rm d}t$ and  $\P$.
We must write a  closure relation between these variables,
for instance by expressing how
    $\P$ depends  on the current state of elastic deformation
and on the applied deformation rate:
$\P(U,\dot{\scalardeformation})$.
We use the following three hypotheses leading to eq. \ref{kinematicHeaviside}.

First we  describe an abrupt transition from elastic
to plastic regime,
as could be the case for an ordered foam \cite{prin83}.
To indicate that T1s appear  when  the absolute value of deformation 
$\vert U\vert $ exceeds the
yield deformation $U_Y$,
we introduce the discontinuous  Heaviside function $\cal H$ (which
is zero for negative numbers, and 1 for  numbers greater than or equal to zero).
This hypothesis can  be relaxed in the  section \ref{sec:Hdefinition}, introducing a
more progressive transition.

Secondly, we account for the hysteresis.
Plastic rearrangements occur when the deformation rate
$\dot\scalardeformation$ and  the current deformation $U$
have the same sign, and again we express it using $\cal H$.
Else, the deformation rate results in elastic unloading, and the deformation
gets smaller than the yield deformation.

Thirdly,  we  use the fact that, in a  slowly sheared  motion,
the only relevant time scale to fix the rate
of plastic rearrangements is  $\dot\scalardeformation$.

Eventually, the plasticity equation writes:
  \begin{equation}
      \P=
{\cal H}(\vert U\vert -U_Y)\;{\cal H}(U\dot\scalardeformation) \;
\dot{\scalardeformation} .
\label{kinematicHeaviside}
\end{equation}
Eq. (\ref{kinematicHeaviside}) can be used to close the system of equations.
  Injecting it  in eq. (\ref{def_P}) yields  an evolution equation of $U$ as a function of the
applied shear rate $\dot{\scalardeformation}$:
\begin{equation}
             \frac{{\rm d}U}{{\rm d}t}
=   \dot{\scalardeformation}\;
      \left[1 -
{\cal H}(\vert U\vert -U_Y)\;{\cal H}(U\dot\scalardeformation) \right].
         \label{eq-evol-u-scalar}
\end{equation}

In eq. (\ref{eq-evol-u-scalar}) $U_Y$ appears as the stable value for 
$U$, that is, a fixed point, at least if $\dot{\scalardeformation}> 
0$; else, the stable fixed point is $-U_Y$.

\begin{figure}[hbtp]
\includegraphics[width=6cm]{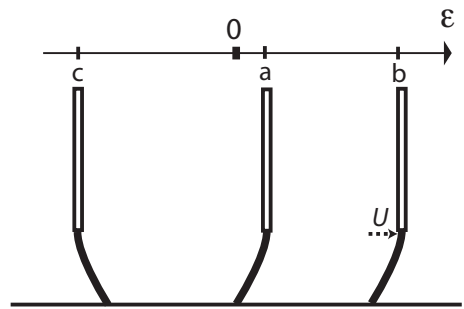}
\caption{Analog scalar system: an elastic brush whose flexion is $U$
stick/slipping on wall. We represent several
states, for an  imposed oscillatory
``painting-like'' motion of the handle $\scalardeformation$, from rest
position 0: (a) onset of sliding to the right, (b)
far-right position, (c) far-left position. }
\label{fig:brush.eps}
\end{figure}

To visualise
    the direction and the
amplitude of the deformation $U$, we suggest an analogy with the 
motion of a  brush on
a wall (Fig. \ref{fig:brush.eps}). The
handle of the brush moves with an oscillatory
position $\scalardeformation$ parallel to the
wall (analog of the imposed scalar deformation of
the material), while the displacement of the
handle with respect to the brush tip is $U$ (the
analog of the internal elasticity of the
material).  The sliding velocity of the contact
point is therefore $\P$ according to equation (\ref{def_P})
and is the analog of plasticity in a material.

\subsubsection{Gradual transition to plasticity}
\label{sec:Hdefinition}

In a disordered foam, for instance with a wide distribution of bubble sizes,
topological rearrangements do not necessarily occur
for the same value of deformation.

We therefore distinguish two different yield deformations.
First, a {\em plasticity yield} $U_{y}$, where deformation ceases to be 
reversible, as defined in material sciences. It is the highest deformation 
for which there is no T1. It is characteristic of the
microstructure, and can
even be close to zero for a very disordered foam.

Second, a {\em saturation yield} $U_{Y}$, the saturation value of elastic deformation at 
which the material can flow with arbitrary large total deformations 
(for instance in Bingham fluids). It is the lowest deformation for which 
the T1s convert the whole total strain into plastic strain. That is, 
$U_{Y}$ is the collapse limit at which a material  structure cannot sustain 
stress. 

We interpolate between $U_{y}$ and $U_Y$ using a function $h(U)$ 
which we call a {\em yield function}.
It should be a growing (or at least non-decreasing) function of
$U$ for $U>0$, and $h(-U)=h(U)$. Moreover, $h(0)=0$, so that
$h(U)\geq 0$ for all $U$. Beside  that, there is no special 
requirement on $h$, which even does
not need to be continuous. Now,  $U_{y}$ is defined as the largest
value of $U$ for which $h(U)=0$, and $U_Y$ as the smallest value of $U$
for which $h(U)= 1$. They do not necessarily correspond to any 
singularity in $h$. We show in Appendix that the precise shape of $h$ 
is unimportant: only $U_{y}$ and $U_Y$ determine material's behaviour. However it is useful for theory to derive analytical equations.

The yield function $h$  depends on
the material under consideration, and can in principle be measured
experimentally.
By definition, $h=0$ corresponds to a purely elastic
state where the elastic deformation follows
the applied deformation.
Conversely, for $h=1$ the
plasticity rate is equal to the
deformation rate.
Such a
    smooth transition from elasticity
to plasticity   generalises
the postulate (\ref{kinematicHeaviside}) as:
\begin{equation}
\P= h(U)\,{\cal
H}(U\dot\scalardeformation)\,
\dot{\scalardeformation}.
\label{def_h}
\end{equation}
Note that we could in principle smoothen out the remaining Heaviside 
function too:
depending on microscopical details, it could be conceivable that some 
T1s appear
during the unloading. We do not explore this possibility here, 
because we seldom observe this effect and it does not seem
to improve significantly the predictions presented below.

Injecting eq. (\ref{def_h}) into   eq. (\ref{def_P}) we obtain:
\begin{equation}
             \frac{{\rm d}U}{{\rm d}t}
=   \dot{\scalardeformation}\;
      \left[1 -
h(U)\,{\cal H}(U\dot\scalardeformation) \right].
         \label{eq-evol-u-smooth}
\end{equation}
Again, the fixed points are $U=\pm U_{Y}$ according to the sign of $\dot\scalardeformation$.

\subsection{Dynamics}

\subsubsection{Slow shear: foam close to equilibrium}

In a foam, bubbles can swap neighbours giving rise to  T1 topological rearrangements.
A T1 is an infinitely short event during which the energy is continuous. Thus it does not dissipate energy by itself, but it brings the foam in
an out-of-equilibrium state. It is thus followed by a dissipation of
energy during the relaxation towards another equilibrium
configuration,
with a microscopical relaxation
time $\tau_{\rm relax}$.

The average life time of a contact between two bubbles is $f^{-1}$, 
where $f$ is the average frequency of T1s per bubble contact. If 
$f\tau_{\rm relax} \ll 1$,
the foam evolves (it is not static)  but
spends most of the
time at or very
close to mechanical equilibrium states. Thus Plateau rules of local 
mechanical equilibrium \cite{wea99}  are
(almost) always
satisfied, up to  corrections  of order $f\tau_{\rm relax} $.

The frequency $f$ can be determined by various causes of 
perturbations, for instance coarsening  \cite{wea99}. In rheology, it 
is determined by the plastic deformation rate $\P$ \cite{mar06}. 
For dimensional reasons,
$f$ is proportional to  $\P$. Since the plasticity amplitude $\P$ is always smaller than the deformation rate $\dot{\scalardeformation}$  (see eqs \ref{kinematicHeaviside} and \ref{def_h}), 
the regime  close to equilibrium is obtained in the slow shear limit:
\begin{equation}
\dot{\scalardeformation}\tau_{\rm relax} \ll 1.
\label{slowshear}
\end{equation}

\subsubsection{Contributions to total stress}

We now include an additional viscous dissipation
from the global deformation of the network of bubbles.
This contribution is not linked to the relaxation of  rearrangements,
and does not modify the slow evolution of deformation. 

We consider two separate contributions to stress, under the 
following hypotheses.
According to experimental tests  \cite{asi03}, we consider an elastic 
contribution to the stress ${\sigma}^\mathrm{el}=\mu {U}$ 
proportional to the
elastic deformation $U$, where $\mu$ is the shear elastic modulus.
It describes a classical elastic behaviour, with a reversible restoring force.

According to the model proposed by Kraynik and co-workers 
\cite{kra87,kra88}, we consider
a viscous
  contribution to the stress due to large scale velocity
gradients:
$\sigma^\mathrm{vis}=\eta\dot{\scalardeformation}$, where  $\eta$ is 
a macroscopic viscosity. It describes a classical fluid behaviour: 
the corresponding
dissipated power is quadratic, proportional to
$\dot{\scalardeformation}^2$.

\begin{figure}[htb]
\begin{picture}(0,0)%
\includegraphics[scale=1]{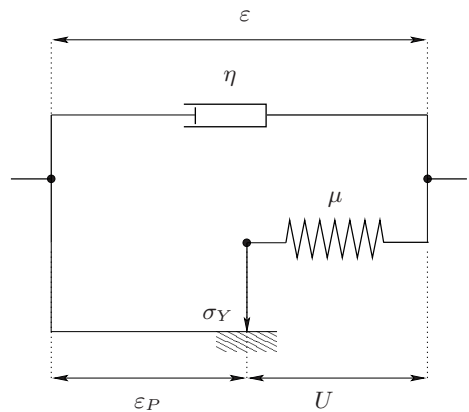}%
\end{picture}%
\setlength{\unitlength}{2072sp}%
\begin{picture}(5804,4896)(3039,-4549)
\put(5941,164){$\varepsilon$}%
\put(4681,-4471){$\varepsilon_P$}%
\put(5500,-3400){$\sigma_Y$}%
\put(6841,-4471){$U$}%
\put(5750,-556){$\eta$}%
%\put(4700,-2000){$\eta_m$}%
\put(7000,-2000){$\mu$}%
\end{picture}%
         \caption{A linear elasto-visco-plastic rheological model.}
         \label{fig-rheology}
\end{figure}

In the spirit of
a polymeric
model \cite{mac94},  we assume that the stresses add up  (Fig. 
\ref{fig-rheology}):
\begin{eqnarray}
{\sigma}&=&\sigma^\mathrm{el}+\sigma^\mathrm{vis}\nonumber\\
&=& \mu {U}+\eta \dot{\scalardeformation}.
\label{eq:stresselasticviscous}
\end{eqnarray}

   The material is characterised
by the coefficients  $\eta$,  $\mu$ and $U_Y$ (and $U_{y}$ in the case of gradual plasticity). 
Measuring experimentally, and understanding
theoretically the physical origin of these
coefficients, requires specific studies for each
material considered: this is beyond the scope of the
present paper.  In principle, they
can  be rank-four tensors (anisotropic material).
They can even vary with the material's state
(non-linear material), for instance in a shear-thinning case.

\begin{figure}
\includegraphics[width=8cm]{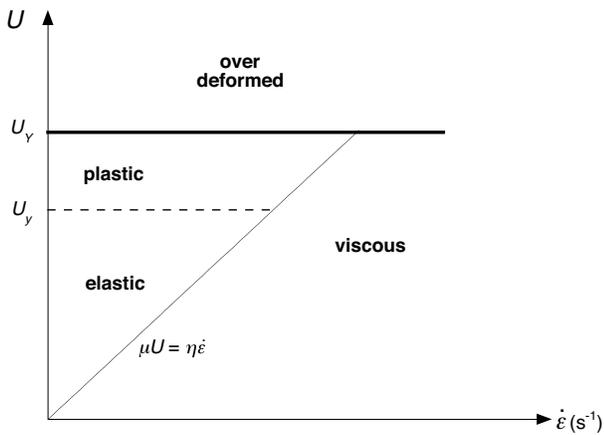}
         \caption{
Scalar   phase diagram for a slowly sheared  foam or an emulsion. Axes are
experimentally measurable  \cite{mar06} local variables:
shear rate $\dot{\scalardeformation}$  and
elastic deformation $U$.
The crossover from elastic to plastic  is defined as the onset
of the first isolated topological rearrangements; it occurs around $U_Y$,
with possible precursors around $U_{y}$. The yield deformation  $U_Y$ 
corresponds to a macroscopic rate of
topological rearrangements.
The crossover
from solid  to fluid  is defined by the equality of viscous
and elastic stresses.
The  slowly sheared regime presented here ceases to be valid when
$\dot{\scalardeformation}$  becomes comparable to  $\tau_{\rm
relax}^{-1}$, inverse
of the microscopical relaxation time.
}
         \label{diagramme_vep}
\end{figure}

  As opposed to the cross-over from elastic to plastic regimes, which 
is topological and is visible on images,
here  the crossover from the elastic to the fluid regime  can be 
detected only by measuring forces.
It
occurs when the viscous contribution to the stress becomes larger
than the elastic one.
Fig. (\ref{diagramme_vep}) thus plots the line corresponding to the crossover:
$
    \mu U = \eta  \dot{\scalardeformation}.
%\label{crossover_detail}
$

Defining    the macroscopic local Weissenberg number  as:
\begin{equation}
\mathrm{Wi}_{M} \equiv \frac{\eta\dot{\scalardeformation}}{\mu},
\label{def_CaM}
\end{equation}
the cross-over between elastic and fluid regime occurs at:
$
\mathrm{Wi}_{M} = U.
$

Upon increasing $\dot{\scalardeformation}$ from the plastic regime where $U$ 
is close to $U_{Y}$, 
we predict a cross-over from a stress bounded by a constant value $\mu U_{Y}$ (with a dissipated power linear in $\dot{\scalardeformation}$, see next section) to a stress
linear in $\eta\dot{\scalardeformation}$ (with a dissipated power quadratic in $\dot{\scalardeformation}$),
characteristic of a viscous friction.
The cross-over from a plastic regime to a fluid regime occurs when $\eta \dot{\scalardeformation}$  is equal to $\mu U_{Y}$, i.e. when $\mathrm{Wi}_{M} = U_{Y}$.

Since the plastic deformation is not bound, in the plastic regime the 
foam can flow indefinitely. As in hydrodynamics, the displacement 
field itself is no longer relevant.
The plastic flow \cite{fra93,fra95} and the viscous flow  \cite{bat00} look 
the same; their difference is not kinematic but dynamic: stresses are 
independent on and proportional to $\dot{\scalardeformation}$, 
respectively.

\subsubsection{Dissipation}

The close to equilibrium criterion (eq. \ref{slowshear}) regards time scales, and is not 
a criterion based
on the absence or presence of dissipation.
Viscous dissipative effects  can indeed be observed when considering 
measurements of the loss
modulus at  very low amplitude oscillations, and hence at very slow shear
rate (as presented below in  Figs.
\ref{fig:GprimeGsecondfigure_general.eps}-\ref{fig:GprimeGsecondRouyer}).
In fact, dissipation  is absolutely
necessary to relax towards equilibrium:
it damps oscillations and decreases
the energy.
Note that a ``quasi-static" regime, that is a succession of 
equilibrium states, necessarily obeys the equilibrium criterion; but it is not 
sure that the reverse is true. In fact, Ref.
\cite{jan06b} claims that in the slowly sheared  Couette flow by \cite{deb01} 
  the velocity profile is determined by the ratio of 
velocity-dependent forces (internal viscosity and external friction on the plates of glass): 
  static simulations are inappropriate.

As already mentioned, a T1 by itself, that is a side swapping,  is an infinitesimally short 
topological event, during which the foam energy is continuous: there 
is no instantaneous dissipation. However, the T1 puts the foam in an 
out-of-equilibrium state. During a time $\tau_{\rm relax}$ the foam 
relaxes to a local energy minimum by dissipating an energy $\delta E$.
A smaller   microscopic dissipation yields a shorter relaxation time 
$\tau_{\rm relax}$, and a larger  {\it instantaneous} dissipated power, 
of order ${\cal P}_{\rm diss} = \delta E/ \tau_{\rm relax}$.
But the amount of energy dissipated, $\delta E$, is independent on 
the dissipation.
Thus the  dissipated power {\it averaged} over a long time (longer 
than  $\dot{\scalardeformation}^{-1}$)  is of order:
\begin{equation}
\left \langle {\cal P}_{\rm diss}\right \rangle
=
f \delta E \sim \P \delta E \sim \dot{\scalardeformation} \delta E.
\label{power}
\end{equation}
This dissipated power is proportional  to $\dot{\scalardeformation}$, and not quadratic as in viscous flows, 
although the microscopical origin is a local viscous dissipation \cite{pug06}.

\begin{figure}[hbtp]
\includegraphics[scale=0.5]{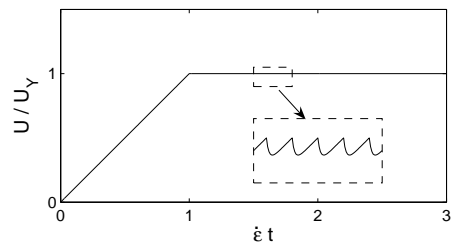}
\caption{Schematic  impact of individual microscopic rearrangements on the
stored elastic deformation $U$, for a constant
loading rate $\dot{\scalardeformation}$.
Rearrangements relax exponentially the
deformation over a time $\tau_{\rm relax}$, 
with here $\dot{\scalardeformation}\tau_{\rm relax}=0.02 \ll 1$.
In the present macroscopic model, rearrangements are coarse-grained. 
}
\label{sawtooth}
\end{figure}

The
elastic deformation is almost independent on the shear rate  $
\dot{\scalardeformation}$ (see figure \ref{sawtooth}).
To obtain a steady shear in a solid regime, when $U$ saturates at the 
value $U_Y$, an experimentalist has to apply an constant external force which 
balances the  average elastic stress, and  does not depend on $\dot{\scalardeformation}$.

Such a dissipated power linear in $\dot{\scalardeformation}$, and a steady 
force which does not depend on $\dot{\scalardeformation}$, are 
characteristic of a solid friction \cite{cou79,bau99}.

\section{Prediction and tests}
\label{predict}

We model the foam response in one type of mechanical experiment, imposed deformation, 
and in two types of rheometrical experiments, creep flow and oscillating shear.

\subsection{Imposed shear}

Here, we calculate the transient response  during a shearing
experiment, that is, the relation
$U(\scalardeformation)$ between applied strain $\scalardeformation$  and
elastic deformation $U$.
For simplicity we take here $\scalardeformation=U=0$ at the start of the
experiment, but that assumption  is   easy to relax.

\begin{figure}[h]
\includegraphics[scale=0.5]{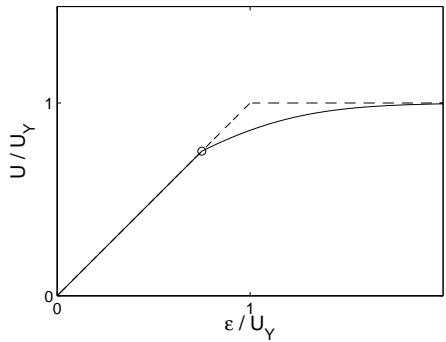}
\caption{Response to imposed shear for two examples of yield functions. 
Dashed solid line: abrupt transition, $h(U)={\cal H}(  U   - U_Y)$ (eq. \ref{kinematicHeaviside}).
Thick line: finite $U_{y}$, 
and linear interpolation
  $h(U)=({  U   -U_{y}})/({U_Y
-U_{y}}){\cal H}(  U   - U_{y})$ (eq. \ref{h_interpolate}). 
Here $U_{y}=0.75\; U_Y$.
See fig. \ref{fig:HandFromrest} for more examples.
}
\label{fig:Fromrest.eps}
\end{figure}

By direct integration, see appendix,
we show the material's response:
the elastic deformation $U$ is close to the imposed
strain $\scalardeformation$ at low applied strain, and
tends to a saturation value at large applied strain.
This robust behaviour does not depend much on the chosen yield function
(see fig. \ref{fig:Fromrest.eps}).

Thus the distribution of bubble sizes
does not affect much the foam's transient response
(as opposed to the liquid fraction, which drastically affects
$U_Y$ \cite{prin83}). This explains  why in the literature the function
$U(\varepsilon)$  is sometimes taken for simplicity
as a piece-wise linear function or as a hyperbolic tangent \cite{jan06}.

This provides
both the physical origin for the
function $\sigma(\varepsilon)$ of the model by Janiaud {\it et al.}
\cite{jan06}, and a justification for their  (up to now arbitrary)
expression $\sigma = \sigma_Y \;
f(\varepsilon/\varepsilon_Y)$: the  function $f$ corresponds
to the present elastic deformation $U$,
while $\varepsilon_Y$ is
the yield deformation they chose equal to 1 for simplification.

\subsection{Creep under constant applied stress}

%\textbf{Comparaison avec \cite{bau04} ??}

\begin{figure}[hbtp]
\def\scaletthisfig{0.4}
{\includegraphics[scale=\scaletthisfig]{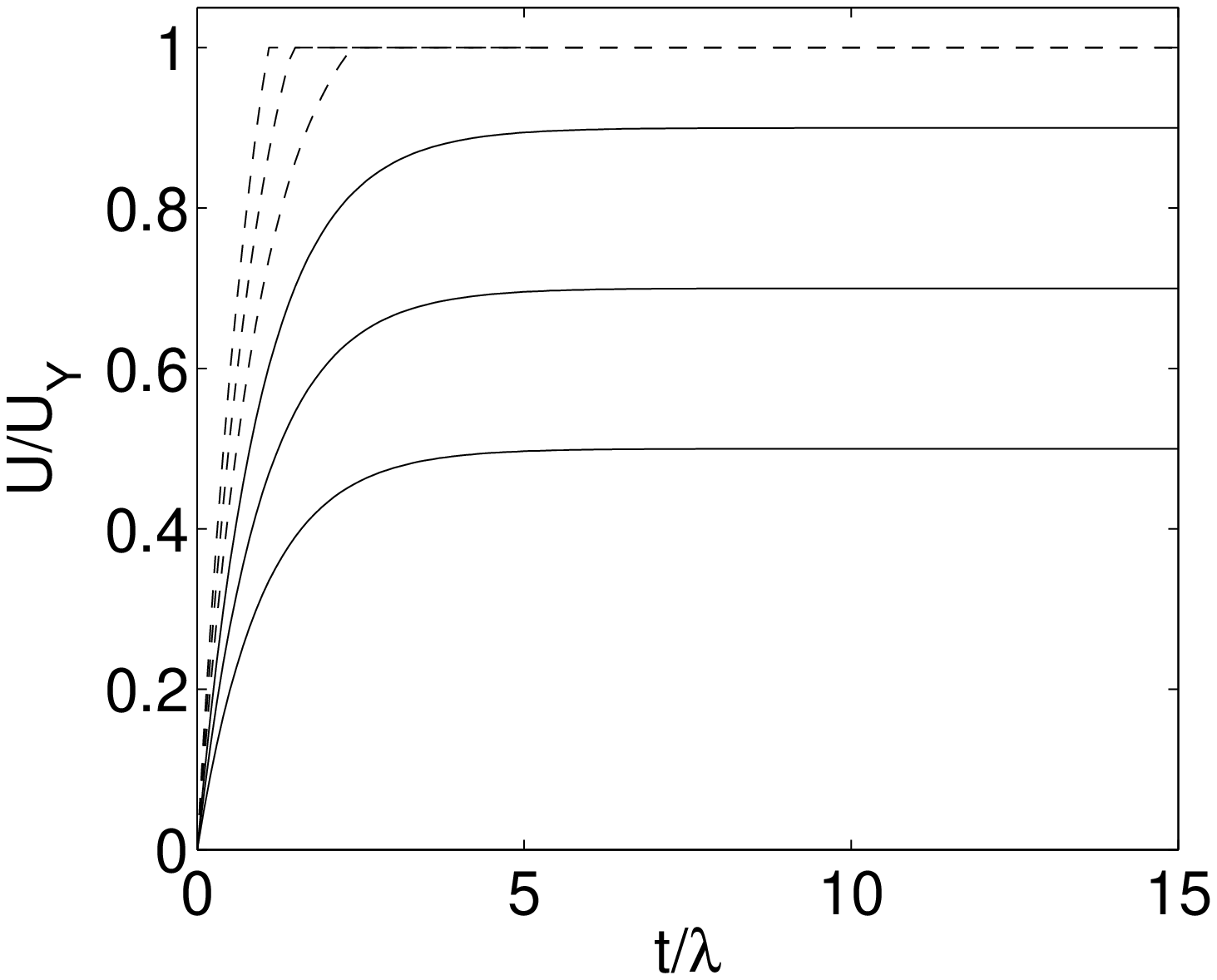}}
{(a)}
{\includegraphics[scale=\scaletthisfig]{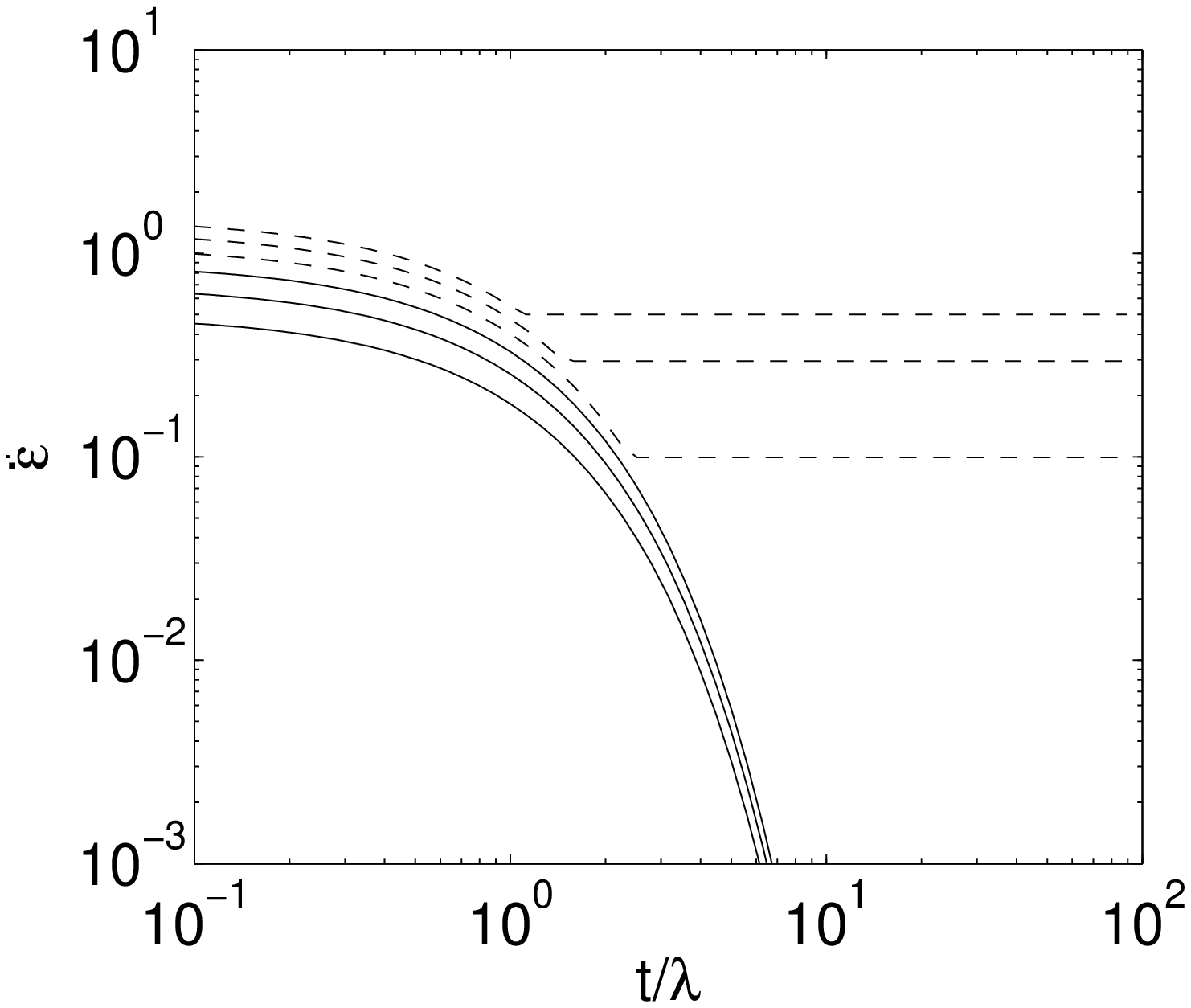}}
{(b)}
\caption{Response to constant applied stress: (a) elastic deformation versus time   and (b) local strain rate versus time, with a sharp transition to plasticity at $U=U_{Y}$.
In both figures the stress  grows from bottom to top:  below yield values, 
$\sigma_\mathrm{app}$ equals to 0.5, 0.7, 0.9 times $\sigma_{Y}$ (solid 
lines) and above yield values,
$\sigma_\mathrm{app}$ equals to 1.1, 1.3, 1.5 times $\mu U_{Y}$ (dashed 
lines). The time is adimensioned by  $\lambda=\eta/\mu$. }
\label{fig:Creep}
\end{figure}

\begin{figure}[hbtp]
\def\scaletthisfig{0.4}
%\grilledix
{\includegraphics[scale=\scaletthisfig]{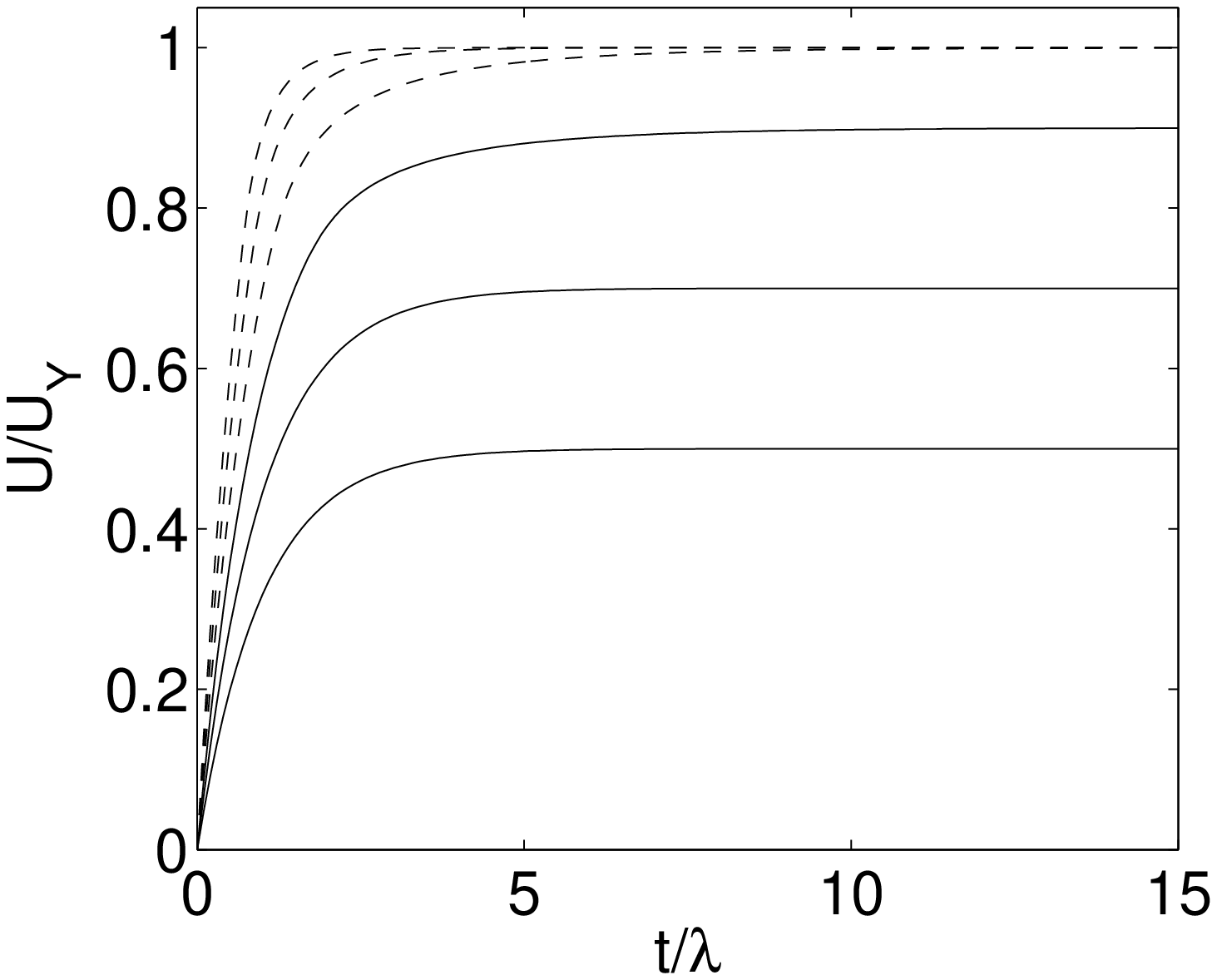}}
{(a)}
{\includegraphics[scale=\scaletthisfig]{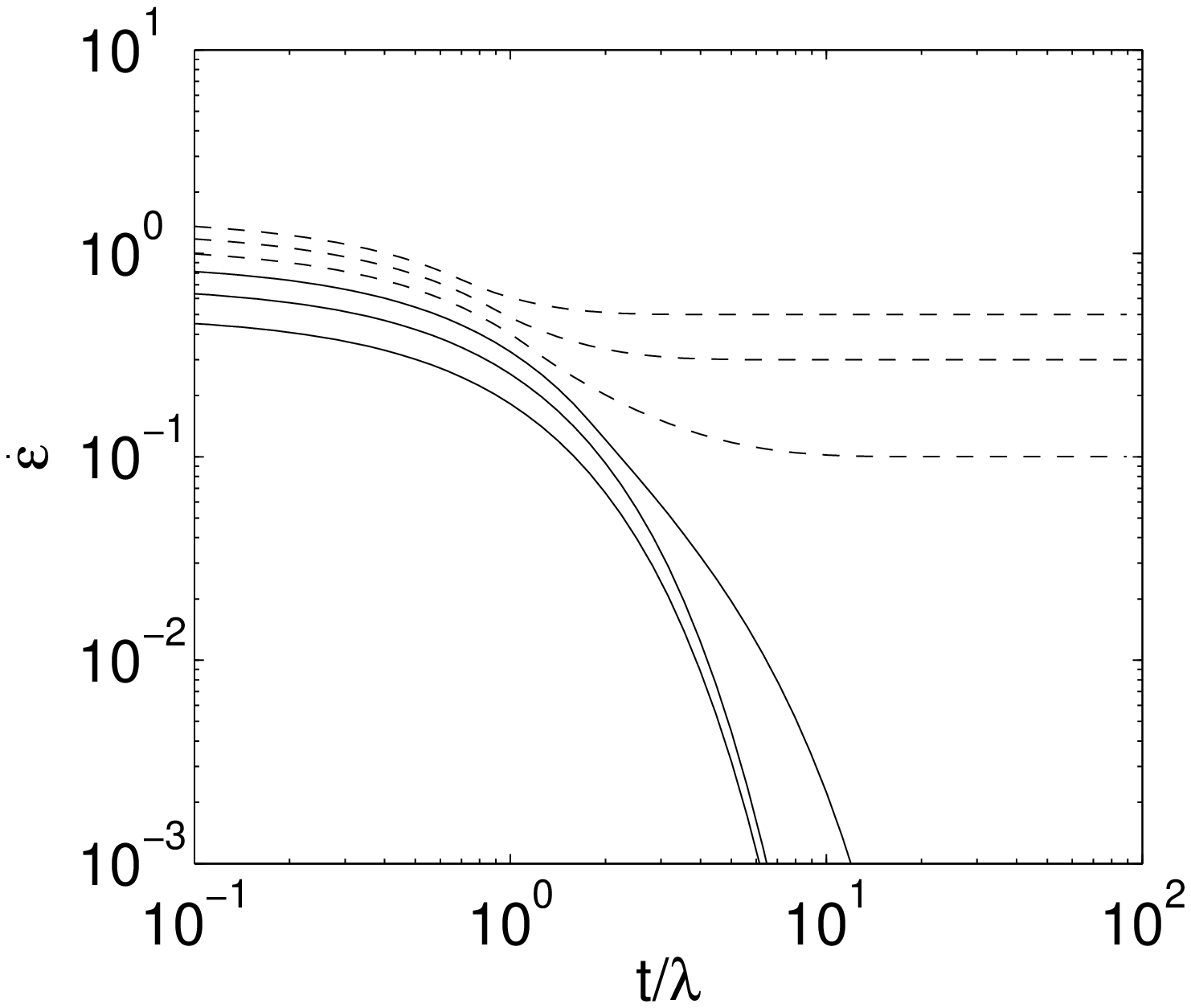}}
{(b)}
\caption{Response to constant applied stress: (a) elastic deformation versus time   and (b) strain rate versus time, with a  smooth appearance of plasticity in between a plasticity onset threshold 
$U_{y}$  and saturation $U_{Y}$. Plasticity starts at $U_{y}=0.75\, U_{Y}$, with the linear interpolation  of eq. \ref{h_interpolate}.
Same legend as previous figure.}
\label{fig:CreepSmooth}
\end{figure}

A creep experiment in a rheometer applies a constant stress
$\sigma_\mathrm{app}$.
It determines the effective viscosity $\eta_\mathrm{eff}$ from the 
steady shear rate: \begin{equation}
\eta_\mathrm{eff}=\lim_{t\rightarrow \infty} \; 
\frac{\sigma_\mathrm{app}}{\dot\varepsilon(t)}.
\end{equation}

The rheological response is found from eq. 
(\ref{eq:stresselasticviscous}) with $\sigma=\sigma_\mathrm{app}$, and from
eq. (\ref{eq-evol-u-scalar}).
The elastic loading and the strain rate are plotted on Fig. (\ref{fig:Creep}).
The elastic deformation saturates to $\sigma_\mathrm{app}/\mu$ when it is 
below the threshold $U_{Y}$, and that it saturates to $U_{Y}$ when 
above the threshold, over  a characteristic time $\lambda=\eta/\mu$.

At long times, the strain rate tends towards vanishing values below 
yield stress (the flow stops), and tends to finite values above the yield: 
 $\dot\varepsilon(t\rightarrow 
\infty)=(\sigma_\mathrm{app}-\mu U_{Y})/\eta$.  We thus deduce the 
effective viscosity:
\begin{eqnarray}
\eta_\mathrm{eff}&=& \infty \quad \mathrm{when} \quad 
\sigma_\mathrm{app}\leq  \mu U_{Y},\\
\eta_\mathrm{eff}&=&\frac{\eta}{1-\frac{\mu 
U_{Y}}{\sigma_\mathrm{app}}}\quad  \mathrm{when} \quad \sigma_\mathrm{app}>\mu 
U_{Y}.
\end{eqnarray}

Taking a smooth plastic transition  (eq. \ref{eq-evol-u-smooth}) does 
not changes the overall features, except that the deceleration times 
below yield are no longer superimposed, see fig. (\ref{fig:CreepSmooth}).

\subsection{Oscillating shear}

\begin{figure}[hbtp]
\includegraphics[scale=0.5]{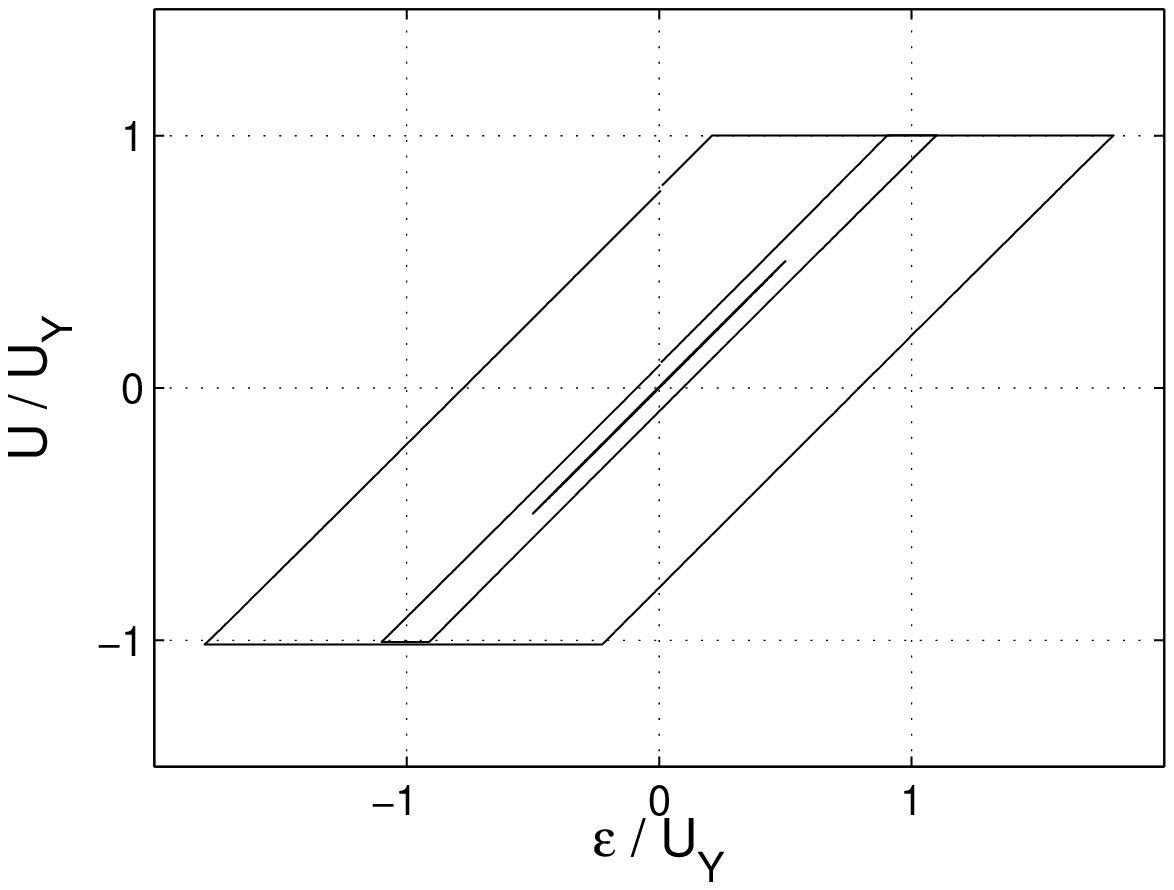}(a)\\
\includegraphics[scale=0.5]{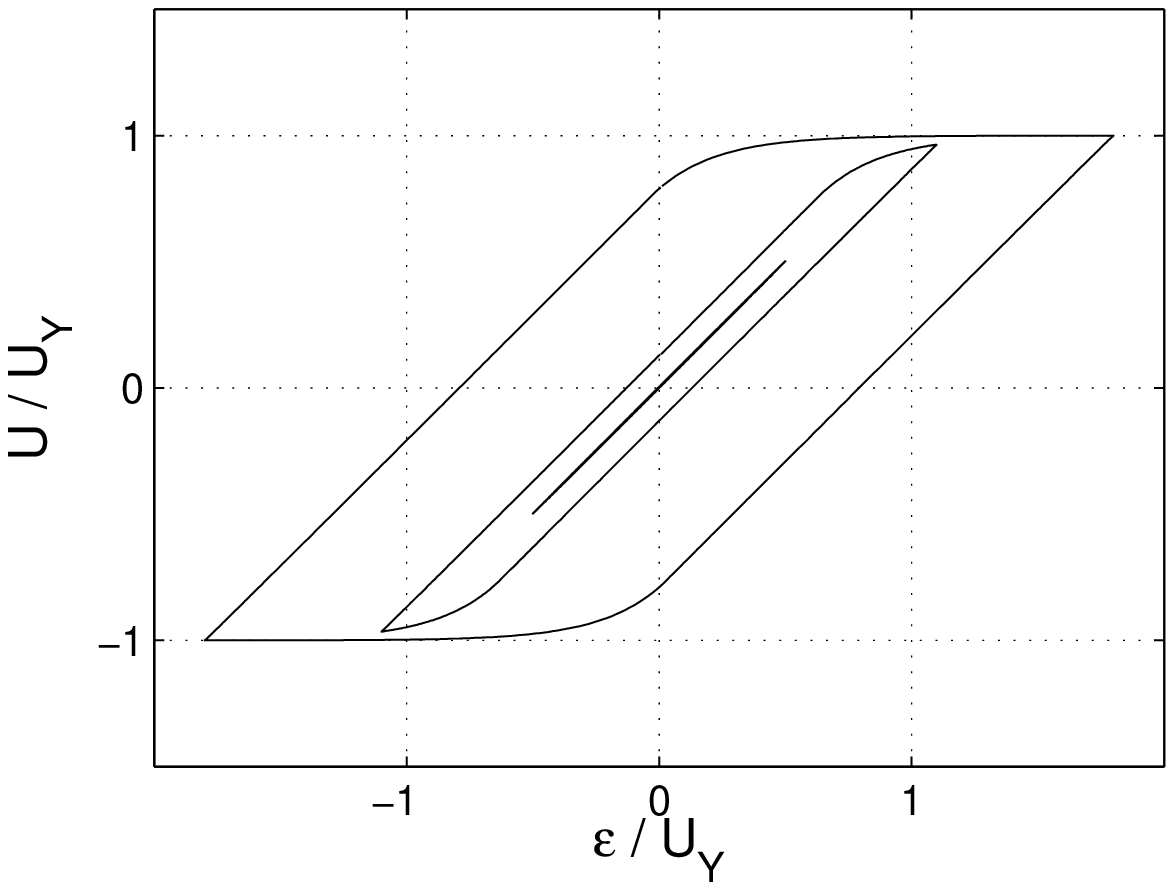}(b)\\
\caption{Long time periodic response to an oscillatory
shear, for three amplitudes: $\varepsilon/U_Y=0.75,
1.1$ and $1.8$. (a) in the case where the plasticity appears 
abruptly at $U_{Y}$, (b) gradual transition in between $U_{y}$ and $U_{Y}$, with $U_{y}/U_{Y}=0.5$ (eq. \ref{h_interpolate}).}
\label{fig:GprimeGsecond_Heaviside.eps}
\label{fig:GprimeGsecond_Square.eps}
\end{figure}

A
Couette apparatus
is another typical rheometry experiment  \cite{mas96,lau02,rou05,hoh05}.
It
   measures the
stress $\sigma(t)$ on the walls while imposing an applied sinusoidal
shear strain of pulsation $\omega=2\pi/T$:
\begin{equation}
\scalardeformation=\gamma \sin(\omega t).
\end{equation}

\subsubsection{Hysteresis cycle and non-linear response}
\label{hyster}

To test the effect of hysteresis of the model we calculate
the response to  such a periodic oscillatory shear
bounded by amplitudes $\gamma$ and $-\gamma$.
In the slow shear  limit, the  frequency does not play any explicit 
role. We can thus keep it fixed without loss of generality.

We integrate 
 eq. (\ref{eq-evol-u-scalar}).
The periodic elastic deformation \textit{vs}
strain curve is plotted on Fig.
(\ref{fig:GprimeGsecond_Heaviside.eps}) top.
The stress response is linear in strain below the
threshold, and saturates above in plastic regime,
exhibiting  a strong hysteresis.

%{\bf (A METTRE EN DISCUSSION ?)}
Reversing the sign of the loading instantly stops
any plasticity and the reponse becomes purely
elastic. Multiple loading does not increase the slope of
the loading part, nor the value of saturation
yield; the foam is described as perfectly plastic.
Such features are observed in experiments on other
amorphous solids \cite{aub99,cou03}
(as opposed to strain-hardening features of
crystalline metals \cite{cha87}).

Integration of eq.
(\ref{def_h})
describes a smooth
variation of
deformation, see Fig. 
(\ref{fig:GprimeGsecond_Square.eps}) bottom.

\subsubsection{Storage and loss moduli: predictions}

In  complex notation the
stress response $\sigma^*$ is linked to the
strain $\scalardeformation^*$ by
$\sigma^*=(G'+iG'')\scalardeformation^*$. Here
$G'$ is  the storage modulus and $G''$ the loss
modulus  of the material,
defined  as
      the in-phase and out phase part of the response
(first term in a Fourier series, see non linear models
\cite{hyu02,sim03,miy06}).

When  increasing the amplitude  $\gamma$ of the imposed
sinusoidal shear strain, the response is first linear  until the
amplitude at which $G'$ and $G''$ start to vary.
      In both the linear and non-linear regimes,
the storage and loss moduli are calculated as:
\begin{eqnarray}
G'&=&-\frac{1}{\gamma^2}\frac{1}{\pi \omega}
\int_0^T \sigma(t)
d\dot{\scalardeformation},
\nonumber\\
G''&=&\quad\frac{1}{\gamma^2}\frac{1}{\pi}
\int_0^T
\sigma(t)d\scalardeformation,\label{eq:G''integral}
\end{eqnarray}
$G'$ is proportional to the area enclosed by the
$(\sigma(t),\dot{\scalardeformation}(t))$ curve,
while $G''$ is proportional to the area enclosed
by the $(\sigma(t), \scalardeformation(t))$ curve.
When  plasticity occurs, the cycle has a
non-vanishing area in the $(\sigma(t),
\scalardeformation(t))$ diagram,  meaning a
non-vanishing loss modulus $G''$.

In the present model $\sigma(t)$  depends on the current elastic
deformation  $U(t)$ and shear rate
$\dot{\scalardeformation}(t)$ (eq. \ref{eq:stresselasticviscous}).
For the case of the abrupt elastic/plastic
transition, the
analytical integration of areas is simple and
provides the following solutions for the moduli. Using
eqs.   (\ref{eq:G''integral}) we obtain, when $\gamma \ll U_Y $:
\begin{eqnarray}
G'&\simeq &\mu
\nonumber\\
G''&=&\eta,
\label{eq:G-low-amplitude}
\end{eqnarray}
 {%}
which is the usual linear visco-elastic regime. Note that our model predicts frequency independant moduli, for a fixed small amplitude $\gamma$.
}
At large amplitudes,  when $\gamma \gg U_Y $:
\begin{eqnarray}
G'&\simeq &\mu
\frac{4}{\pi}\left(\frac{U_Y}{\gamma}\right)^{3/2},
\nonumber\\
G''&=&\mu\frac{4U_Y}{\pi\gamma}+\eta. \label{eq:Gsecondhigh}
\end{eqnarray}

%{\bf (A METTRE EN DISCUSSION ?)}
These asymptotic dependencies in $\gamma^{-3/2}$ and  $\gamma^{-1}$, 
respectively, are obtained analytically and are robust with respect 
to the model. The analytical expression of $G'$ and $G''$ over the 
whole range of $\gamma$, but  with $\eta=0$ is explicitly presented 
in refs. \cite{lab04,hoh06}.

\begin{figure}
\includegraphics[width=8.5cm]{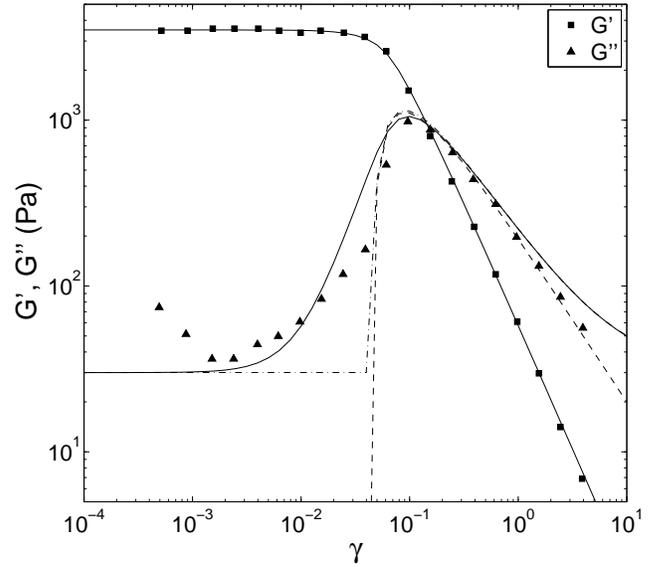}
\caption{
Storage and loss moduli {\it versus}  strain amplitude for a monodisperse emulsion.
Symbols: experimental $G'$ (circles) and $G''$ (triangles) in
   a close-packed
emulsion (Fig. 1  of ref.
\cite{mas95}, fraction of the continuous phase 20\%,
droplet size 0.53 $\mu$m, oscillation pulsation  $\omega=1$
rad $s^{-1}$).
Lines:   models for $G'$ (solid line), and for $G''$ with  an abrupt
transition
(dashed line), with viscosity (dash-dotted line), with viscosity and a smooth yield function
$h=(U/U_Y)^2$, $U_{y}=0$ (solid line).
Model parameters:
shear modulus $\mu=1.7\; 10^3$ Pa, yield deformation $U_Y=0.045$,
viscosity $\eta=30$ Pa.s.
}
\label{fig:GprimeGsecondfigure_general.eps}
\end{figure}

\begin{figure}
\includegraphics[width=8.5cm]{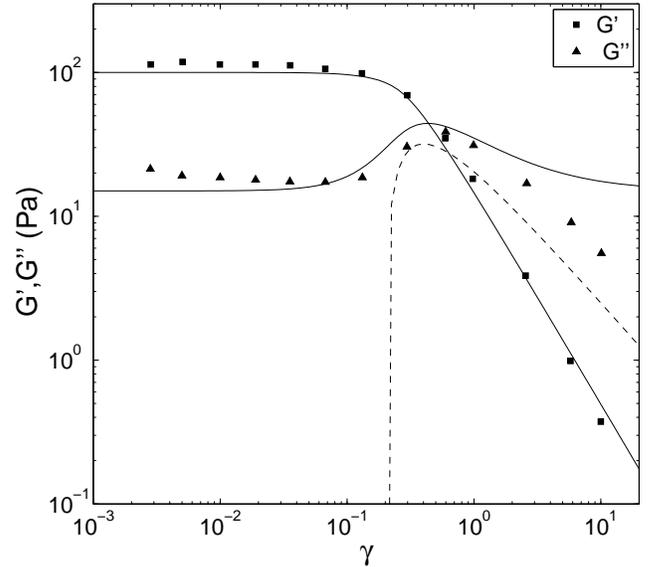}
\caption{
Same as Fig. (\ref{fig:GprimeGsecondfigure_general.eps}) for a
polydisperse foam
\cite{sai99}.
Liquid fraction 5\%, bubble  size 40 to 70 $\mu$m,
$\omega=1$
rad $s^{-1}$.
Lines:   models for $G'$ (solid line), and for $G''$ with  an abrupt
transition
(dashed line),  with viscosity and a smooth yield function
$h=(U/U_Y)^2$, $U_{y}=0$ (solid line).
Model parameters:
$\mu=100$ Pa,  $U_Y=0.2$,
 $\eta=15$ Pa.s.
}
\label{fig:GprimeGsecondJalmes}
\end{figure}

\begin{figure}
\setlength{\unitlength}{1cm}
\begin{picture}(8,8)(0,0)
\put(-0.5,0){\includegraphics[width=9cm]{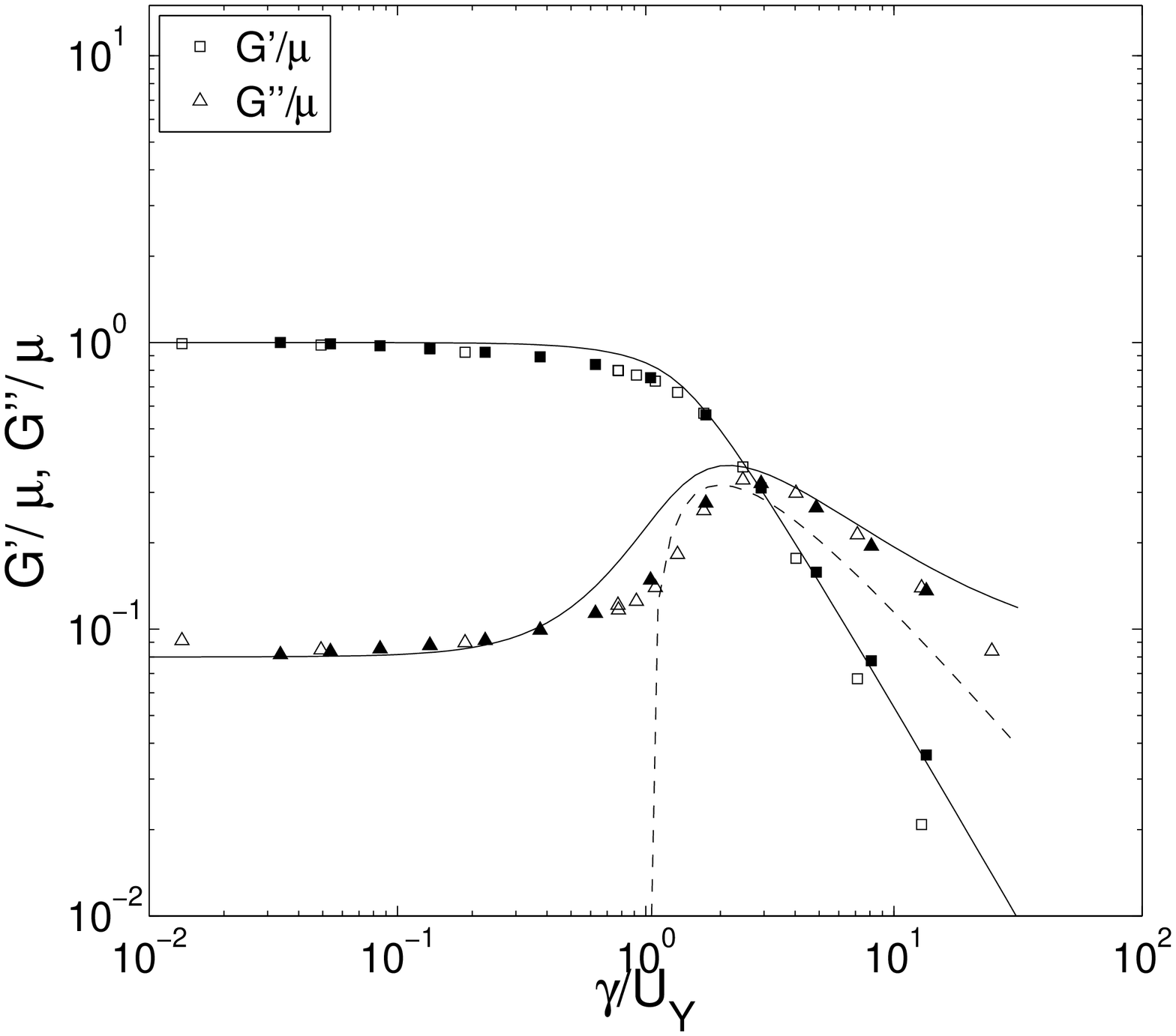}}
\put(4.7,4.5){\includegraphics[width=3.5cm]{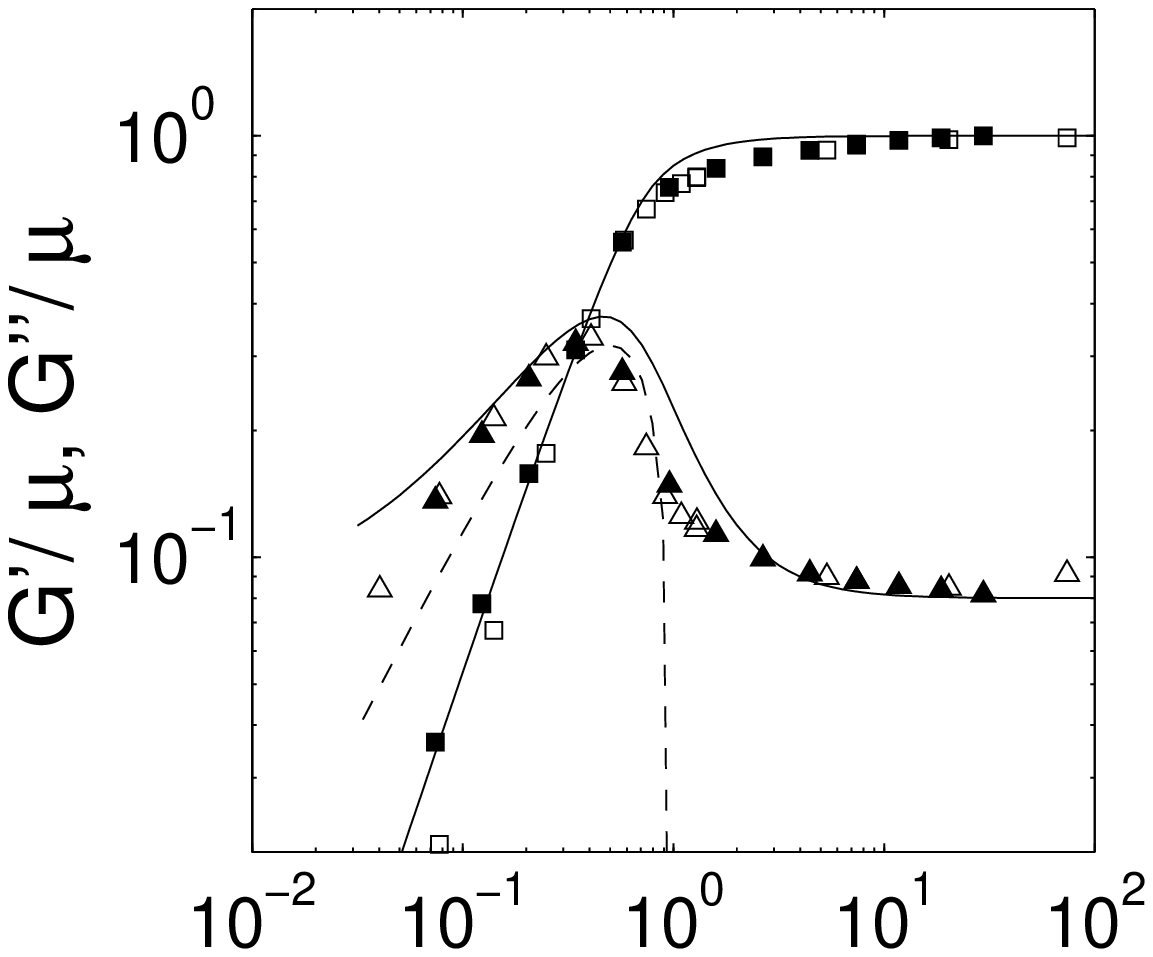}}
\put(6,4.5){$\omega/(\dot\gamma_{0}/U_{Y})$}
\end{picture}
\caption{
Same as Fig. (\ref{fig:GprimeGsecondfigure_general.eps}) for a
monodisperse foam \cite{rou05}.
Liquid fraction 8\%, bubble  size 21 $\mu$m,
$\omega=1$
rad $s^{-1}$.
Data of $G'$ and $G''$ are normalised by $\mu$ and $\gamma$ by $U_Y$,
while  $\eta \omega / \mu=0.08$.
Same legend as figure \ref{fig:GprimeGsecondJalmes}. 
 {%
Inset: Strain-rate frequency superposition: same data plotted as a function of pulsation $\omega$, for a given maximum strain rate $\dot\gamma_{0}=\omega \gamma$ \cite{wys06}.} 
}
\label{fig:GprimeGsecondRouyer}
\end{figure}

For a smooth yield function $h$, 
predictions are obtained numerically and plotted on Figs.
(\ref{fig:GprimeGsecondfigure_general.eps}-\ref{fig:GprimeGsecondRouyer}).

%Programme Matlab GprimeGsecond_Mason2.m

%Programme Matlab GprimeGsecond_Jalmes95_SVM.m

%Programme Matlab GprimeGsecond_Rouyer.m

\subsubsection{Comparison with experiments  on  emulsions and foams}

%{\bf (
%Pour ces 3 figures de $G'$ et $G"$ (et si possible aussi les 
%precedentes) il faudrait utiliser une legende coherente:  la meme 
%pour chaque figure. D'ailleurs, le choix des lignes pleins ou 
%pointillees en fonction du message qu'on veut faire passer. Par 
%ailleurs, ces figures seraient peut-etre plus adaptees au message du 
%texte si on utilisait la forme avec $U_{y}$ plutot que $(U/U_Y)^2$.
%Enfin, pourquoi ne pas montrer $G'$ pour chaque modele?
%D'ou la suggestion, pour toutes ces figures:
%1) ligne de points pour $G',G"$ avec modele EP abrupt.
%2) ligne de traits pour $G',G"$ avec modele VEP abrupt.
%3) ligne continue pour $G',G"$ avec modele VEP lisse avec $U_{y}$.
%)}

Rheometry
measurements of  monodisperse
emulsions \cite{mas95}
(Fig. \ref{fig:GprimeGsecondfigure_general.eps})
and
polydisperse foams  \cite{sai99,rou05}
(Figs. \ref{fig:GprimeGsecondJalmes} and \ref{fig:GprimeGsecondRouyer})
directly
yield, without hypotheses, the values
of the material parameters required by the model.
The shear modulus $\mu$ is read from  the value of $G'$
at low amplitude.
The viscosity  $\eta$ is read from  the value of $G''$
at low amplitude (or the value of the minimum, in Fig.
\ref{fig:GprimeGsecondfigure_general.eps}, 
where the two data points at $\gamma<10^{-3}$ have too large error bars to be taken into account, according to T. Mason, private communication).
The yield deformation $U_Y$ is read from the
intersection of low amplitude plateau of $G'$ and its large amplitude
$-3/2$ exponent power-law.

A purely elasto-plastic model is enough to predict $G'$ correctly,
over the whole range of amplitude,
including the $-3/2$ exponent power-law.
This simplest model also describes  correctly the
large amplitude trend for $G''$.

The  low amplitude value of $G''$
can be modelled by including a viscosity (dash-dotted line of figure  \ref{fig:GprimeGsecondfigure_general.eps}),
which confirms that viscosity is relevant even in such  slowly 
sheared models.
This procedure is at the expense of a slight
over-prediction at large amplitudes.
This latter aspect suggests a possible shear-thinning, that is a
decrease of the viscosity $\eta$
with  the shear rate, similarly to the observed reduction of the
drag  of foams
in motion in channels \cite{prin83}.

The agreement between the data and the model, without adjustable
parameter, is good.
It is still improved, even for $G''$ at intermediate
amplitudes, if we account for the fact that the first T1s appear gradually
at a value $U_{y}$ lower than $U_Y$ (solid lines on figures for $G''$).
Even the value of $U_{y}$ itself is not very important, and in order to avoid introducing a free parameter  we use here $U_{y}=0$ and a smooth yield function.

 {%
These result suggest that data can be rescaled with the yield deformation $U_{Y}$, and it suggest a rescaling when plotting data as a function of frequency, following the strain-rate frequency superposition method (SRFS \cite{wys06}). 
This method consider measurements with a fixed
maximum strain rate $\dot\gamma_{0}=\omega \gamma$, it is therefore equivalent to vary frequency or oscillation amplitude. The natural rescaling  for pulsation 
that appears is 
$\omega/b(\dot\gamma_{0})=\omega/(\dot\gamma_{0}/U_{Y})=U_{Y}/\gamma$,
see  inset of Fig (\ref{fig:GprimeGsecondRouyer}), and our model predicts the global shapes of the moduli curves as observed in  \cite{wys06}.
The characteristic frequency $b(\dot\gamma_{0})$ is here linear in $\dot\gamma_{0}$, also in agreement with the trend observed in \cite{wys06} for large enough strain rates.
}

\section{Discussion}
\label{disc}

\subsection{Discussion of the predictions}

Independently  from us, H\"ohler et al.
solve a purely elasto-plastic model \cite{hoh06,lab04}. Since they neglect the viscosity,
they can eliminate $U$  
and replace it by $\sigma/\mu$.
The agreement of their model with data of  Fig.  \ref{fig:GprimeGsecondRouyer}, as 
well as other experimental data, is good for  $G'$
over the whole range of amplitude; and also for $G''$ but only at large 
amplitude, where the dissipation comes from the relaxation after T1s
rather than from $\eta$.

Our models describes better: (i) $G"$ 
at low amplitude using viscosity; and 
(ii) $G''$ at intermediate
amplitudes, if we account for the fact that the first T1s appear
at a value $U_{y}$ lower than $U_Y$.

The predicted curves are robust with respect to $U_{y}$. This implies 
that we do not
need to fit it; but
that, conversely, we are not yet able to deduce $U_{y}$ from $G''$ data.
If we had
a direct experimental measurement of $U_{y}$, we could inject it in the
model to predict $G''$
     at intermediate amplitudes, near $U_Y$, but the resulting
predictions would
be very similar to the present ones.

While $U_{y}$ corresponds to the onset of isolated plastic events,
at a deformation $U_Y$ the plastic events have a macroscopic effect:
they catch the total strain; the foam flows without increasing its 
deformation any longer.

The effective viscosity $\eta_\mathrm{eff}$ diverges to infinite values when 
$\sigma_\mathrm{app}/\mu \rightarrow U_{Y}^+$. This means  that the foam 
comes close to its yield deformation, in the fluid sense. This change in 
behaviour was shown by \cite{cru02}, and modelled by a granular model 
with a velocity dependant friction. Here, the dynamics is entirely 
driven by a constitutive equation with shear-rate  independent 
parameters. It is the transient elastic loading that drives a 
transient flow, which stops when the stress is not strong enough.

The trends of figure (\ref{fig:Creep}) agree  with experimental data on 
various material reported in \cite{cru02}, at least with granular 
materials and emulsion. The present model does not predict the 
apparent shear-thinning behaviour observed with their experimental 
data with foams \cite{cru02}, where an increase of shear rate with 
time is found \cite{cou02}.
 {%
The low amplitude predictions for the visco-elastic regime (see Eq. \ref{eq:G-low-amplitude}) are frequency independant moduli. The limitation of the present model is thus that it does not fully capture the slight increase  of the loss moduli at low frequency, and the $\omega^{1/2}$ trend at large frequencies (see for instance the case of foams \cite{hoh05}).
}

\subsection{Elastic, plastic, viscous  model}

%\subsubsection{Phase diagram}

A complete model for an elastic, plastic, viscous foam requires to 
recognise the role of three physical variables $U$, 
$\dot{\scalardeformation}$, $\P$. There is a relation between them 
(eq. \ref{def_P}).
Unless specific approximations apply,  a foam's representative volume 
element (RVE) is characterised by two independent variables: we 
suggest to select the local elastic deformation $U$,
and the local shear rate  $\dot{\scalardeformation}$, which are 
intuitive and physically relevant.
Both of course depends on the
sample's past history, but this history plays no explicit role. Both
are always defined,
whether in elastic, plastic or viscous regime \cite{aub03}.
Two recent works \cite{wys06,mar06z} find that $G'$ and $G''$
depend on the strain amplitude and on the strain rate (rather than on 
the frequency),
in the same spirit as our phase diagram (fig. \ref{diagramme_vep}).

Each volume element can thus be plotted as a point in a phase diagram
(Fig. \ref{diagramme_vep}); that is, the
    $(\dot{\scalardeformation},U)$ plane \cite{pic05}.
In a heterogeneous flow, different volume elements of the same foam
are plotted as different points. A volume element's evolution is a
trajectory on this plane.
{\it Simple} materials correspond to the axes of the plane:
pure elastic and pure plastic regimes on the vertical axis,
pure viscous regime (Navier-Stokes) on the horizontal axis.

%\subsubsection{Elastic strain}

The importance of $U$ is the most original feature of the present model:
$U$ cannot be entirely determined by
$\dot{\scalardeformation}$  since the latter
can change sign; $U$ cannot be entirely determined by $\sigma$ if the 
viscous contribution
is not negligible.

The yield function describing the occurrence of plasticity can be linked to the traditional  hardening modulus, used for the description of plastic materials \cite{sim98}.  It is defined as $K=d\sigma/d\varepsilon_{p}$, while the elastic modulus is $\mu=d\sigma/U$. In the present model, the hardening modulus is dependant on the elastic deformation: $K=1/h(U)-1$.
It therefore vanishes when $U$ tends to its saturation value $U_{Y}$: at this point the material does not harden any more.

A deformation  beyond $U_Y$ is not accessible when starting from rest
(Fig. \ref{diagramme_vep}). But the foam could initially be
prepared (for instance artificially \cite{eli99}) in a
configuration very far from equilibrium.
Under a
    steady shear rate
$\dot{\scalardeformation}$,  the deformation $U$ always tends towards
$U_Y (\dot{\scalardeformation})$, whether from below or from above.

%\subsubsection{Slow shear regime}

In a flowing foam, there is always a viscous dissipation. 
Its contribution becomes dominant in front of the plastic dissipation if $\dot{\scalardeformation} > \mu U / \eta$.
This is compatible with the slow shear criterion, 
$\dot{\scalardeformation} \ll \tau_{\rm relax} ^{-1}$  (ie on a second Weissenberg number that is here $\mathrm{Wi}_{m}=\dot{\scalardeformation} \tau_{\rm relax}\ll 1$), if there is a 
scale separation between the microscopic relaxation time $\tau_{\rm relax}$
  towards local equilibrium and the large scale time $\eta/\mu$.
The dimensionless ratio  $ \mu\tau_{\rm relax} / \eta$ of 
microscopic to macroscopic times is analogous to the parameter $I$ of 
granular materials \cite{jop06}.

\subsection{Perspectives}
\label{sec:Perspectives}

This model could in principle be generalised to  higher velocity 
gradients \cite{sar06}. This would require a high flow 
velocity varying over a small scale, and
  $\tau_{\rm relax} $ could play an explicit role. The deviation from 
equilibrium, of order
$\dot{\scalardeformation}  \tau_{\rm relax} $, would become 
significant: for instance,
under a steady shear the limit value of $U$ could become larger than $U_Y$.

A rheometer such as a Couette apparatus  can measure 
$\sigma_{12}=\sigma$ (tangential force per unit wall surface) and 
$\varepsilon_{12}=\varepsilon/2$ (components of the symmetrized 
deformation gradient), in a coordinate system aligned with walls.
 {%
For comparison with tensorial data it is especially important to bear in mind that there is a {\em factor $1/2$}: the threshold $U_{Y}$ on oscillation amplitude $\varepsilon$ as measured by a Couette rheometer, corresponds to a threshold $U_{Y}/2$ on the tensorial deformation $\varepsilon_{12}$. 
}
The present  scalar approach can  be generalised to take into 
account such an influence of the orientation of material deformation, 
as well as spatial variations
\cite{mar06,rau06testexp}.

%\subsection{Foam as a continuous material}

The present paper is a contribution to a lively  debate. Can a foam be 
described as a continuous material? We tend here to answer ``yes", in 
the same spirit as many recent papers
which describe or predict  rheological properties at large scale
\cite{asi03,dollet_local, mar06,jan06,lab04,hoh06,sar06}.
 {%
Statistical  descriptions of fluctuations and their correlations
\cite{sol97,fal98,pic05,cou05,cru02} are then useful in describing even 
more complex rheological behaviour such as shear banding \cite{kab03} or  growing correlation lengths cale near the glass transition \cite{Berthier2005}.
}
%thus appear as refinements to correct the average description.
Interestingly, even in granular materials, where very large scale 
fluctuations are known to occur, a recent paper emphasises the 
dominant role of the ``continuous material" description based on 
averages \cite{jop06}.

\section*{Acknowledgments}

We  thank E. Janiaud for critical reading of the manuscript, R. 
H\"ohler for comparison of our calculations before
publication,
F. Rouyer for providing experimental data,
S. Ataei Talebi,
I. Cheddadi,
B. Dollet,
C. Quilliet,
C. Raufaste,
and
P. Saramito
for
discussions,  T. Mason and A. Saint-Jalmes for comments on their experimental data. 

\appendix
\section*{Appendix: Elastic-plastic transition}
\label{sec:appendix}

\subsection{Transient response from rest}

We assume (in this section only) that {\it the deformation rate 
$\dot{\scalardeformation}$ keeps a constant sign}.
Under this essential assumption,
we can calculate analytically the transient response  during a shearing
experiment. That is, the relation
$U(\scalardeformation)$ between applied strain $\scalardeformation = 
\int  \dot{\scalardeformation}\; {\rm d}t$ and
elastic deformation $U$.

The yield function $h$ is defined to interpolate between $h(U)= 0$ 
for $0<U<U_{y}$, and $h(U_Y)=1$.
By direct integration, eq.
(\ref{eq-evol-u-smooth}) yields:
\begin{equation}
\scalardeformation = \int_0^U \; \frac{{\mathrm d}U}{1-h(U)}.
\label{gamma_(U)}
\end{equation}
Here, without further loss of generality, we have also  assumed  (but 
it is  easy to relax) that  $\scalardeformation=U=0$ at the start of 
the
experiment, and that $\dot{\scalardeformation}\geq 0$, so that $U \geq0$ too.
%%%%%%{\bf (J'ai donc vire les valeurs absolues)}

Eq. (\ref{gamma_(U)})  yields the function
$\scalardeformation(U)$, which can be inverted
to obtain $U(\scalardeformation)$. These
functions can be  measured on experiments
and compared with
predictions derived from direct  measurements of $h(U)$.

Whatever the function $h(U)$, eq. (\ref{gamma_(U)})
implies that $U  \approx \scalardeformation$  as long as $U< U_{y}$:
applied and elastic deformation are equal in  the elastic
regime. At the onset of plasticity  (or
topological changes),  $U>U_{y}$, they differ.
When $U$ gets close to  $U_Y$ the r.h.s. of eq. (\ref{gamma_(U)}) 
diverges. Thus,
when $\scalardeformation$ increases arbitrarily,
$U$ tends asymptotically towards
the saturation value $U_Y$.

\subsection{Examples of yield functions $h$}

\begin{table}[htbp]
       \centering
       \begin{tabular}{@{} ll @{}}
         \toprule
         Yield function $h $& Elastic response $U(\scalardeformation)/U_Y$ \\
         \hline
           $ {\cal H}(  U   - U_Y)$  &
$\scalardeformation- {\cal
H}(\scalardeformation/U_Y-1)\scalardeformation$ \\
finite $U_{y}$ (eq. \ref{h_interpolate})
\quad &
Eq. (\ref{finite_Um})
\\
$(   U /U_Y)^0$& $0$ \\
            $(   U /U_Y)^1$ & $1-\exp (-\scalardeformation/U_Y)$ \\
           $(   U /U_Y)^2$ & $\tanh (\scalardeformation/U_Y)$ \\
         $\sin^2(   U /U_Y)$ & $\arctan (\scalardeformation/U_Y)$ \\
         \hline
       \end{tabular}
       \caption{Elastic deformation for different examples of  yield
function $h$, for a non-deformed initial
condition $U(0)=0$ and  with
$\dot\scalardeformation$ of constant sign.}
       \label{tab:hfunctions}
\end{table}

\begin{figure}[h]
\includegraphics[scale=0.5]{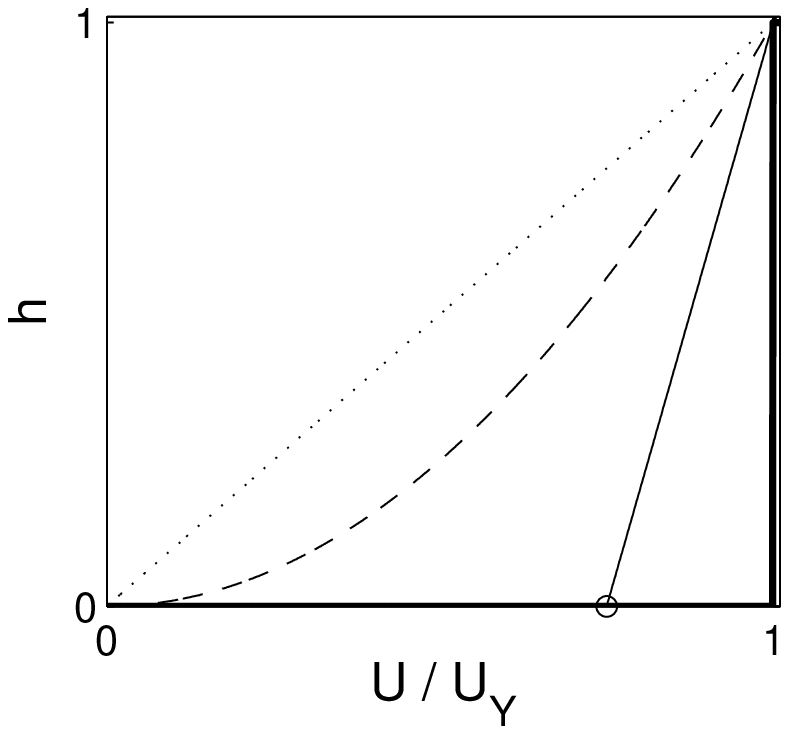}
(a)\\
\includegraphics[scale=0.5]{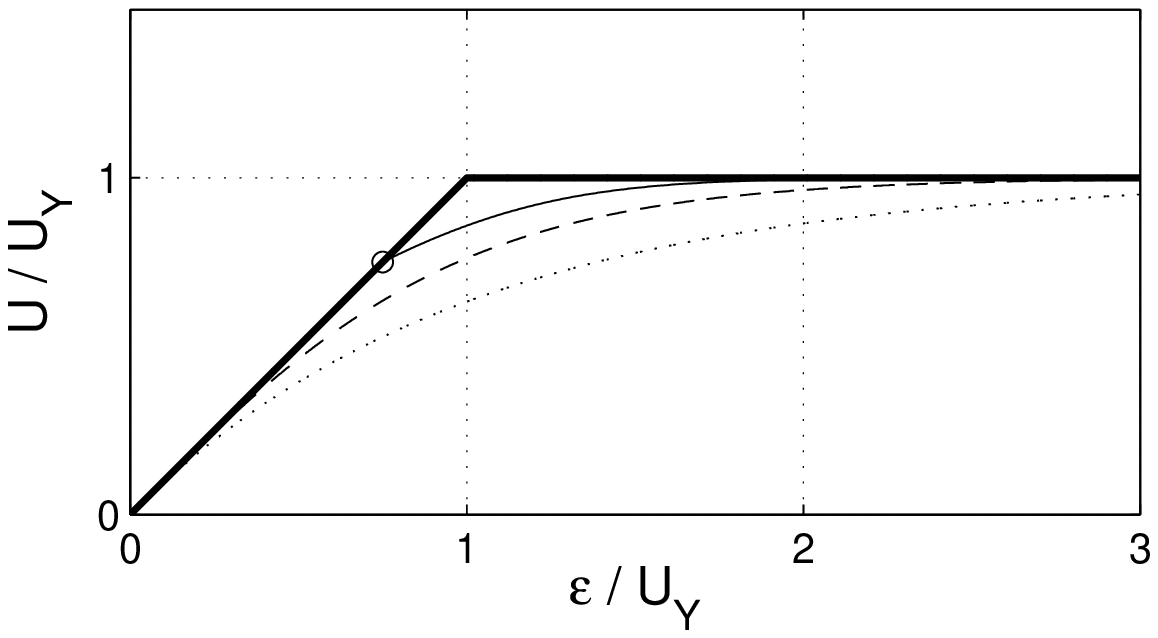}
(b)
\caption{
%{\bf
%(Tracer en abscisse jusqu'a la valeur 3; choisir les lignes en 
%coherence avec les figures de $G'$, $G"$; donc peut-etre supprimer le 
%cas   ``vanishing $U_{y}$ and linear interpolation,
%$h=(U/U_Y)$"
%  qui n'est pas utilise dans le reste de l'article; ou au contraire 
%decider de tracer toutes celles qui sont dans la table? faire une 
%figure (a) qui trace les fonctions $h(U)$ correspondantes, qui sont 
%dans la table I, et mettre la figure actuelle en (b))}
Responses from rest
for some examples of yield functions in Table I.
(a) $h(U)$; (b) $U/U_Y$ versus $\scalardeformation/U_Y$: since 
they are very similar,
for clarity only some of them  are plotted.
Thick solid line: abrupt transition, $h(U)={\cal H}(  U   - U_Y)$.
Thin line: finite $U_{y}$, here $U_{y}=0.75\; U_Y$,
and linear interpolation
  $h(U)=({  U   -U_{y}})/({U_Y
-U_{y}}){\cal H}(  U   - U_{y})$ (eq. \ref{h_interpolate}).
Dashes: vanishing $U_{y}$ and quadratic interpolation,
$h=(U/U_Y)^2$.
Dots: vanishing $U_{y}$ and linear interpolation,
$h=(U/U_Y)$.
}
\label{fig:HandFromrest}
\end{figure}

Table \ref{tab:hfunctions} proposes a few examples of yield functions $h$, and some are plotted on figure \ref{fig:HandFromrest}.

Eq.  (\ref{eq-evol-u-scalar}) is only a particular case of the more general eq.
(\ref{eq-evol-u-smooth}),
with $h$ being the discontinuous Heaviside function:
\begin{equation}
h(U)={\cal H}(  U   - U_Y).
\label{eq:hHeaviside}
\end{equation}
Eq. (\ref{gamma_(U)})
  thus includes  the case of  the abrupt transition.

    An example of a yield function with finite $U_{y}$ is
a piece-wise linear  function:
\begin{eqnarray}
   U   \leq U_{y} \quad & \quad  h(U) = 0, \nonumber \\
  U  \geq U_{y} \quad & \quad   h(U) = \frac{  U  -U_{y}}{U_Y -U_{y}}.
\label{h_interpolate}
\end{eqnarray}
and
eq. (\ref{gamma_(U)}) yields directly:
\begin{eqnarray}
\label{finite_Um}
\scalardeformation \leq U_{y} \quad  \quad
&U(\scalardeformation)  & =  \scalardeformation
, \nonumber \\
\scalardeformation\geq U_{y} \quad  \quad
& U(\scalardeformation) & =  \frac{\scriptstyle U_Y}{\scriptstyle U_{Y}-U_{y}} 
 \nonumber \\
 &  & - \left(\frac{ \scriptstyle U_Y}{\scriptstyle U_{Y}-U_{y}}-\scriptstyle U_{y}\right)\; e^{-\left(
\frac{ \scalardeformation -  U_{y}   }{U_Y -U_{y}}
\right)}
.
\end{eqnarray}

We can interpolate between abrupt and smooth
transitions, using the family of  model power-law
yield functions:
\begin{equation}
h(U)=\left(\frac{  U   }{U_Y}\right)^n.
\label{eq:hpowerlaw}
\end{equation}
For instance,  the quadratic expression
$h(U)=(U/U_Y)^2$ yields $U(\scalardeformation)=
U_Y\tanh(\scalardeformation/U_Y)$.
With these functions,
plasticity appears more or less gradually, as
soon as $  U >0$.
That is, $U_{y}=0$.
The limit  $n\rightarrow \infty$ is the
   Heaviside function (eq. \ref{eq:hHeaviside}).

More generally, the yield
function can be thought as the  convolution of
the Heaviside function ${\cal H}$  and a distribution of yield values $p_Y$:
\begin{equation}
h(U)=\int p_Y(U_Y) {\cal H}(\vert U\vert-U_Y) dU_{Y}.
\label{eq:hconvolution}
\end{equation}
 For instance, if the distribution of yield values  $p$ 
is a Dirac peak at
$U_Y$, it results in a Heaviside   yield function $h$
(eq. \ref{eq:hHeaviside}).

\subsection{Robustness with respect to the choice of $h$}

Some functions $U(\scalardeformation)$
from table \ref{tab:hfunctions}
are plotted on figure \ref{fig:HandFromrest}b.
Strikingly, they do  not depend much on the actual expression of
$h(U)$. In fact, only the expression of $h$ near $U_Y$ matters; the
relation between
$\scalardeformation$ and $U$ is robust.
The elastic deformation $U$ is close to the imposed
strain $\scalardeformation$ at low applied strain, and
tends to a saturation value at large applied strain.

%In a foam, the liquid fraction changes $U_Y$ \cite{prin83}: this 
%drastically affects
%the foam's transient response.
%The  disorder of bubble sizes, which tends to decrease $U_{y}$, has 
%less effect; while the actual details of the bubble size 
%distribution, which determines the expression of $h$, has almost no 
%effect (Table I and Fig. \ref{fig:Fromrest.eps}).
%This explains  why   for simplicity the function
%$U(\scalardeformation)$ can indifferently be taken
%as a piecewise linear function or as a hyperbolic tangent \cite{jan06}.

The only important feature of $h$ is its derivative $h'$ just below 
the yield point:
\begin{equation}
h' = \left( \frac{dh(U)}{dU} \right)_{U \to U_Y^-}.
\end{equation}
It determines how the fraction in the r.h.s. of eq. (\ref{gamma_(U)}) diverges.
Thus  $U(\scalardeformation)$ is not the same if $h'$ is zero or 
infinite, or even not defined as in eq. \eqref{eq:hHeaviside}.
If it is infinite, $U$ reaches the saturation value at a finite value of  applied  deformation.

Conversely, if $h'$ is finite,  as in most examples of Table I,
the behaviour is universal.
In eq. (\ref{gamma_(U)}), the fraction diverges as $(U_Y-U) h'$. 
Thus,  whatever the  value of $h'$,
$\scalardeformation(U)$ diverges logarithmically and
$U(\scalardeformation)$ tends exponentially towards $U_Y$.

%\subsection{Shear thinning to account for negative slope of $G"(\gamma)$}
%To add ??

%\bibliography{/Users/Philippe/Library/texmf/bibtex/Bibliographie/foams,
%/Users/Philippe/Library/texmf/bibtex/Bibliographie/plasticity}

\end{document}